# Diurnal variations in the stratosphere of the ultrahot giant exoplanet WASP-121b

Thomas Mikal-Evans[1,2] *, David K. Sing[3,4], Joanna K. Barstow[5], Tiffany Kataria[6], Jayesh Goyal[7,8], Nikole Lewis[8], Jake Taylor[9,10], Nathan. J. Mayne[11], Tansu Daylan[2,12], Hannah R. Wakeford[13], Mark S. Marley[14], Jessica J. Spake[15]

[1]Max Planck Institute for Astronomy, Heidelberg, Germany. [2]Department of Physics and Kavli Institute for Astrophysics and Space Research, Massachusetts Institute of Technology, Cambridge, MA, USA. [3]Department of Earth & Planetary Sciences, Johns Hopkins University, Baltimore, MD, USA. [4]Department of Physics & Astronomy, Johns Hopkins University, Baltimore, MD, USA. [5]School of Physical Sciences, The Open University, Milton Keynes, UK. [6]Jet Propulsion Laboratory, California Institute of Technology, Pasadena, CA, USA. [7]National Institute of Science Education and Research (NISER), Homi Bhabha National Institute, Jatni, Odisha, India. [8]Department of Astronomy and Carl Sagan Institute, Cornell University, Ithaca, NY, USA. [9]Institute for Research on Exoplanets, Département de Physique, Université de Montréal, Montréal, Canada. [10]Department of Physics (Atmospheric, Oceanic and Planetary Physics), University of Oxford, Oxford, UK. [11]Physics and Astronomy, College of Engineering, Mathematics and Physical Sciences, University of Exeter, Exeter, UK. [12]Department of Astrophysical Sciences, Princeton University, Princeton, NJ, USA. [13]School of Physics, University of Bristol, Bristol, UK. [14]Lunar and Planetary Laboratory, Department of Planetary Sciences, University of Arizona, Tucson, AZ, USA. [15]Division of Geological and Planetary Sciences, California Institute of Technology, Pasadena, CA, USA.

* e-mail: tmevans@mpia.de

**The temperature profile of a planetary atmosphere is a key diagnostic of radiative and dynamical processes governing the absorption, redistribution, and emission of energy. Observations have revealed dayside stratospheres that either cool[1,2] or warm[3,4] with altitude for a small number of gas giant exoplanets, while other dayside stratospheres are consistent with constant temperatures.[5-7] Here we report spectroscopic phase curve measurements for the gas giant WASP-121b,[8] which constrain stratospheric temperatures throughout the diurnal cycle. Variations measured for a water vapour spectral feature reveal a temperature profile that transitions from warming with altitude on the dayside hemisphere to cooling with altitude on the nightside hemisphere. The data are well explained by models assuming chemical equilibrium, with water molecules thermally dissociating at low pressures on the dayside and recombining on the nightside.[9,10] Nightside temperatures are low enough for perovskite ($CaTiO_3$) to condense, which could deplete titanium from the gas phase[11,12] and explain recent non-detections at the day-night terminator.[13-16] Nightside temperatures are also consistent with the condensation of refractory species such as magnesium, iron, and vanadium. Detections[15-18] of these metals at the day-night terminator suggest, however, that if they do form nightside clouds, cold trapping does not efficiently remove them from the upper atmosphere. Horizontal winds and vertical mixing could keep these refractory condensates aloft in the upper atmosphere of the nightside hemisphere until they are recirculated to the hotter dayside hemisphere and vaporised.**

    WASP-121b is an ultrahot (>2,000 Kelvin) gas giant exoplanet orbiting an F6V star every 30.6 hours.[8] Previous observations have shown that the dayside hemisphere of WASP-121b has a thermal inversion, with a temperature profile that increases with increasing altitude or, equivalently, with decreasing atmospheric pressure.[3,19] The thermal inversion is thought to be caused by the presence of optical absorbers capturing a substantial fraction of incident stellar radiation at low pressures in the atmosphere.[20,21] Observations of the planet during transit geometry have identified a number of such absorbers, including gaseous Fe, Mg, Cr, V, and VO.[14-18]

Two full-orbit phase curves of WASP-121b were observed at epochs in 2018 and 2019 with the Hubble Space Telescope (HST) Wide Field Camera 3 (WFC3) infrared spectrograph. For each observation, a time series of spectra was acquired using the G141 grism, which covers the 1.12-1.64μm wavelength range. Further technical details of the observations are provided in Methods. A broadband light curve was produced by summing each spectrum in the time series across the full wavelength range (Extended Data Fig. 1). This light curve was fit by simultaneously modelling the planet signal and instrumental systematics, as described in Methods. Quantitative results are reported in Extended Data Fig. 2 and the best-fit model is shown in Fig. 1a, with orbital phases of 0 and 0.5 corresponding to the primary transit and secondary eclipse mid-times, respectively. As described in Methods, the best-fit phase curve model was inverted to generate a global temperature map for WASP-121b (Fig. 1b). On the dayside hemisphere temperatures exceed 3,000 Kelvin and drop to below 1,500 Kelvin in the coolest regions of the nightside hemisphere.

To recover the planetary emission spectrum at different orbital phases, light curves were generated for twelve spectroscopic channels across the 1.12-1.64μm wavelength range (Extended Data Fig. 3). These light curves were analysed using a similar method to the broadband light curve fit (see Methods). The measured emission maxima of the spectroscopic phase curves give the spectrum of the planetary dayside hemisphere, shown in Fig. 2a. In addition, phase-resolved emission spectra were generated by averaging the planetary flux inferred from the spectroscopic light curve fits across sixteen bins in orbital phase. The planetary emission spectrum recovered immediately prior to the primary transit is shown in Fig. 2b and is comprised almost entirely of emission from the nightside hemisphere of the planet. At intermediate phases, the emission received from WASP-121b emanates from a combination of the dayside and nightside hemispheres.[22-24]

Wavelengths covered by the data are sensitive to an opacity band of $H_2O$ vapour and continuum opacity of $H^-$ (Fig. 2). The measured shape and amplitude of these spectral features allow the chemical abundances and vertical temperature profile of the atmosphere to be inferred.[1-3] To recover these properties from the data, a retrieval analysis was first performed on the dayside emission spectrum. As described in Methods, the overall heavy element enrichment ('metallicity') of the atmosphere was allowed to vary, with the relative abundances of individual elements held fixed to solar ratios, and a one-dimensional analytic temperature profile was adopted with three free parameters. As the metallicity and temperature profile were varied in the fitting, chemical abundances were computed assuming chemical equilibrium. Results of this analysis are reported in Extended Data Fig. 4, including a measured metallicity of $[M/H] = 0.76^{+0.30}_{-0.62}$ (approximately 1-10x solar). A second retrieval analysis was also performed for the nightside emission spectrum. Due to the lower signal-to-noise, the atmospheric metallicity was held fixed to the value determined from the dayside retrieval analysis, leaving only the three temperature profile parameters free. As described in Methods, the contribution to the overall emission from the narrow crescent of dayside hemisphere visible at this phase was also factored in to the modelled emission.

The inferred dayside and nightside emission spectra are shown in Fig. 2 and the corresponding pressure-dependent temperatures, $H_2O$ abundances, $H^-$ abundances, and contribution functions are shown in Fig. 3. A dayside thermal inversion is inferred at the pressures probed by the data (below ~30 millibar), consistent with previous results.[3,19] On the dayside, the $H_2O$ abundance drops sharply with decreasing pressure, due to thermal dissociation of molecules.[9,10] Thermal ionisation also raises the abundance of free electrons, which bind with atomic hydrogen to form $H^-$ (refs 6, 7, 9, 10, 25). As temperatures decrease on the nightside, $H_2O$ molecules recombine at low pressures. Rotational-vibrational transitions of $H_2O$ molecules at near-infrared wavelengths increase the efficiency of radiative cooling in the upper atmosphere (Extended Data Fig. 5), resulting in temperature profiles that cool with decreasing pressure on the nightside (Fig. 3a). As described in Methods, consistent results for the dayside and nightside hemisphere properties were obtained when retrievals were performed at intermediate phases (Extended Data Figs 6-9) and when the assumption of chemical equilibrium was relaxed (Fig. 2 and Extended Data Fig 10).

These measurements provide empirical constraints for the theory that refractory species may be lost from the upper atmosphere of highly-irradiated planets due to cold trap processes.[11,12] For example, due to the large temperature contrasts expected between the dayside and nightside hemispheres, refractory species could condense on the nightside and settle to deeper layers of the atmosphere, despite dayside temperatures being high enough to maintain them in the gas phase. However, day-night cold trapping of this kind might be avoided if vertical mixing is vigorous within the atmosphere, allowing condensates to be suspended aloft long enough for lateral winds to return them to the dayside hemisphere.[26,27] Alternatively, condensates may gravitationally settle to deeper layers of the atmosphere and subsequently re-enter the gas phase as they are returned to lower pressures by updrafts.[12]

Condensation curves for relevant refractory species[27-29] are shown in Fig. 3a, namely, corundum ($Al_2O_3$), perovskite ($CaTiO_3$), VO, Fe, forsterite ($Mg_2SiO_4$), and enstatite ($MgSiO_3$). The corundum, perovskite, and Fe condensation curves are crossed during the WASP-121b diurnal cycle (Fig. 3a) and it is also likely that those of forsterite, VO, and enstatite are crossed in the coolest regions of the nightside hemisphere (Fig. 1b). It is particularly notable that temperatures drop low enough for Fe, Ca, Mg, and V to condense, as recent observations have revealed these heavy metals in the gas phase at the day-night terminator.[15-18] Vertical mixing must therefore be operating efficiently within the atmosphere of WASP-121b, to avoid day-night cold trapping. This also appears to be the case for another ultrahot gas giant, WASP-76b, for which gaseous Fe has been detected at the eastern terminator but not detected at the cooler western terminator, where it has presumably condensed.[30] However, non-detections of Ti and TiO at the day-night terminator of WASP-121b complicate this picture,[13-16] as these gases should also form condensates such as perovskite and $TiO_2$ on the nightside.[27-29] It would be surprising if Ti-bearing condensates are efficiently cold trapped while other refractory species avoid a similar fate. This is especially true for V, which is chemically similar to Ti but an order of magnitude less abundant in the solar neighbourhood.[31] For now, this remains an outstanding puzzle, with a solution that may depend on additional factors such as variations in surface energies between different condensate species.[32]

The dayside and nightside emission spectra predicted by a cloud-free three-dimensional general circulation model (GCM) simulation generated for this study (Methods) and results from two published GCMs[9] are shown in Fig. 2. Good agreement with the data is obtained, suggesting that the GCMs have successfully captured much of the interplay between the radiation, chemistry, and dynamics of the WASP-121b atmosphere. The broadband phase curve predicted by the GCM simulation run for this study is also shown in Fig. 1, having an overall amplitude in respectable agreement with the data. However, around the quadrature phases (i.e. 0.25 and 0.75), the GCM underpredicts the planetary emission (see also Extended Data Figs 3 and 6). Nightside clouds are unlikely to explain this discrepancy, as they would be expected to lower the emission by blocking radiation from deeper, warmer layers of the atmosphere. Refractory clouds forming close to the terminator region, however, could potentially boost the emission received from the dayside crescent by reflecting light from the host star.[19,33] Another possible explanation may be provided by the optically thick exosphere of WASP-121b that has been observed to extend to the planet's Roche limit,[17] well below the pressure range considered by the GCMs. Heated layers of the stellar-facing exosphere would be maximally visible at quadrature, raising the overall emission received from the planet, whereas at superior and inferior conjunction, the data are sensitive to deeper atmospheric layers due to the zenith viewing geometry (Fig. 3d), and as such are well matched by the GCM predictions (Fig. 2). Furthermore, the GCMs did not include opacities for gaseous metals such as Fe and Mg, which are known to be present in the atmosphere of WASP-121b[15-18] and could contribute substantially to the outgoing emission.[10] These effects, along with others not considered here, such as latent heat release from the dissociation/recombination of hydrogen[25] and atmospheric drag,[7,34] should be investigated in future modelling.

The dynamics and chemistry of ultrahot gas giants such as WASP-121b are exotic by solar system standards, driven by dramatic contrasts in the irradiation environments of the dayside and nightside hemispheres. Until now, it has proven challenging to explore these diurnal variations due to the narrow infrared wavelength coverage of HST.[7,35] For WASP-121b, these wavelengths are fortuitously sensitive to a pressure range that allows the transition from inverted to non-inverted temperature profiles to be mapped globally. Further insights are anticipated with the *James Webb Space Telescope*, which will enable higher signal-to-noise spectroscopy across the broader 0.8-11μm wavelength range. This will provide fuller coverage of the $H^-$ opacity continuum and access to stronger $H_2O$ bands at longer wavelengths, breaking the degeneracy between the two species. Additional spectral features, such as the CO spectral band at 4.5μm, will provide further leverage for constraining the chemical composition, thermal structure, and wind patterns of the atmosphere.

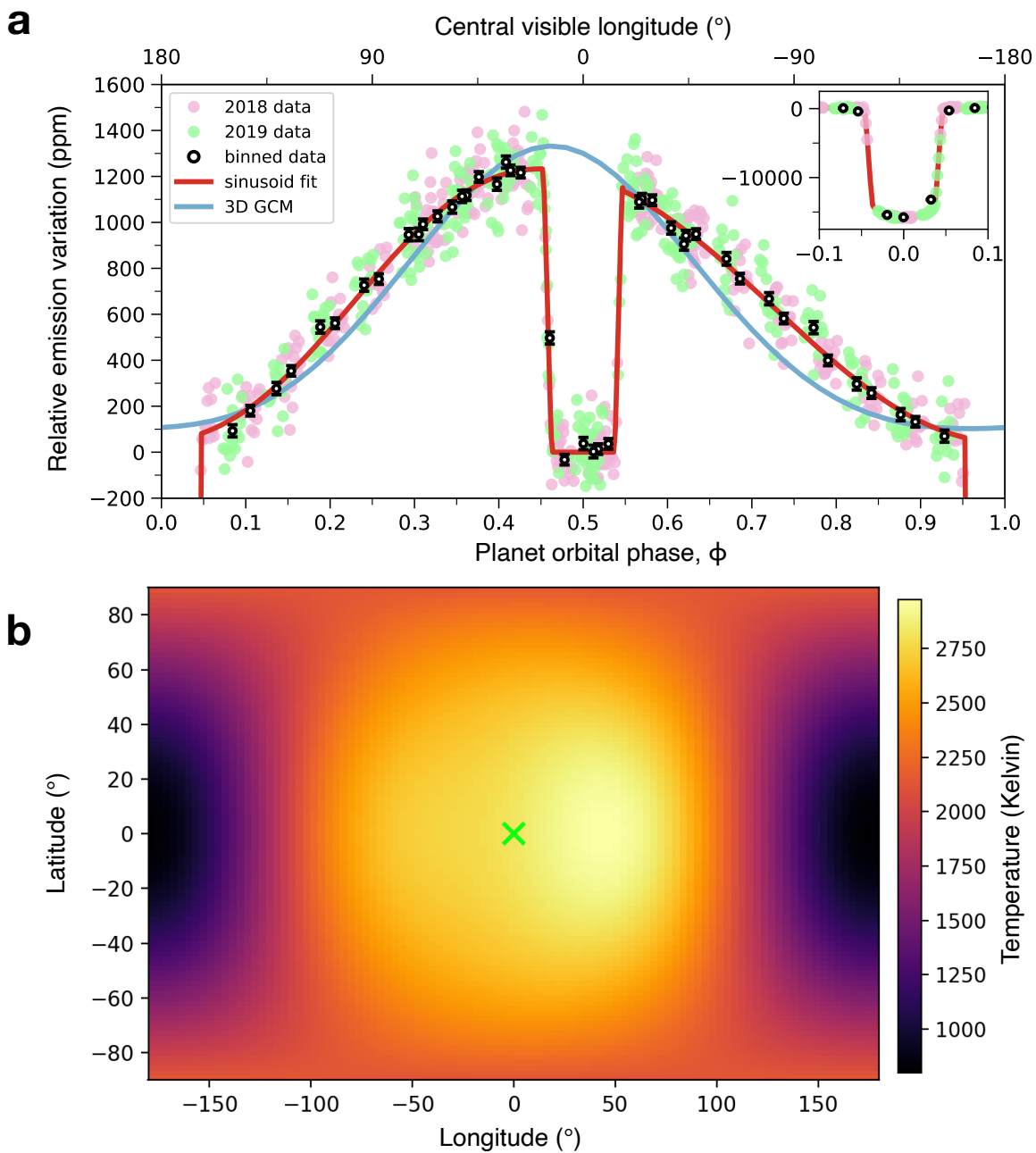

**Fig. 1. Broadband phase curve and inverted temperature map for WASP-121b. a,** Planet emission relative to the host star emission as a function of the planetary orbital phase. Pink and green circles show WFC3 measurements made at epochs in 2018 and 2019, respectively. Black circles show binned data for individual HST orbits, with error bars indicating the 1σ measurement uncertainties. Red line shows the maximum likelihood second-order sinusoidal model, including primary transit and secondary eclipse signals. Blue line shows the prediction of a 3D GCM simulation. Inset shows the full primary transit signal and has the same units as the main axes. **b,** Latitude-longitude temperature map obtained by inverting the maximum likelihood phase curve model as described in Methods. Note that this is a non-unique inversion and it assumes that the temperature map can be described by a low-order spherical harmonics expansion. Green cross indicates the location of the substellar point.

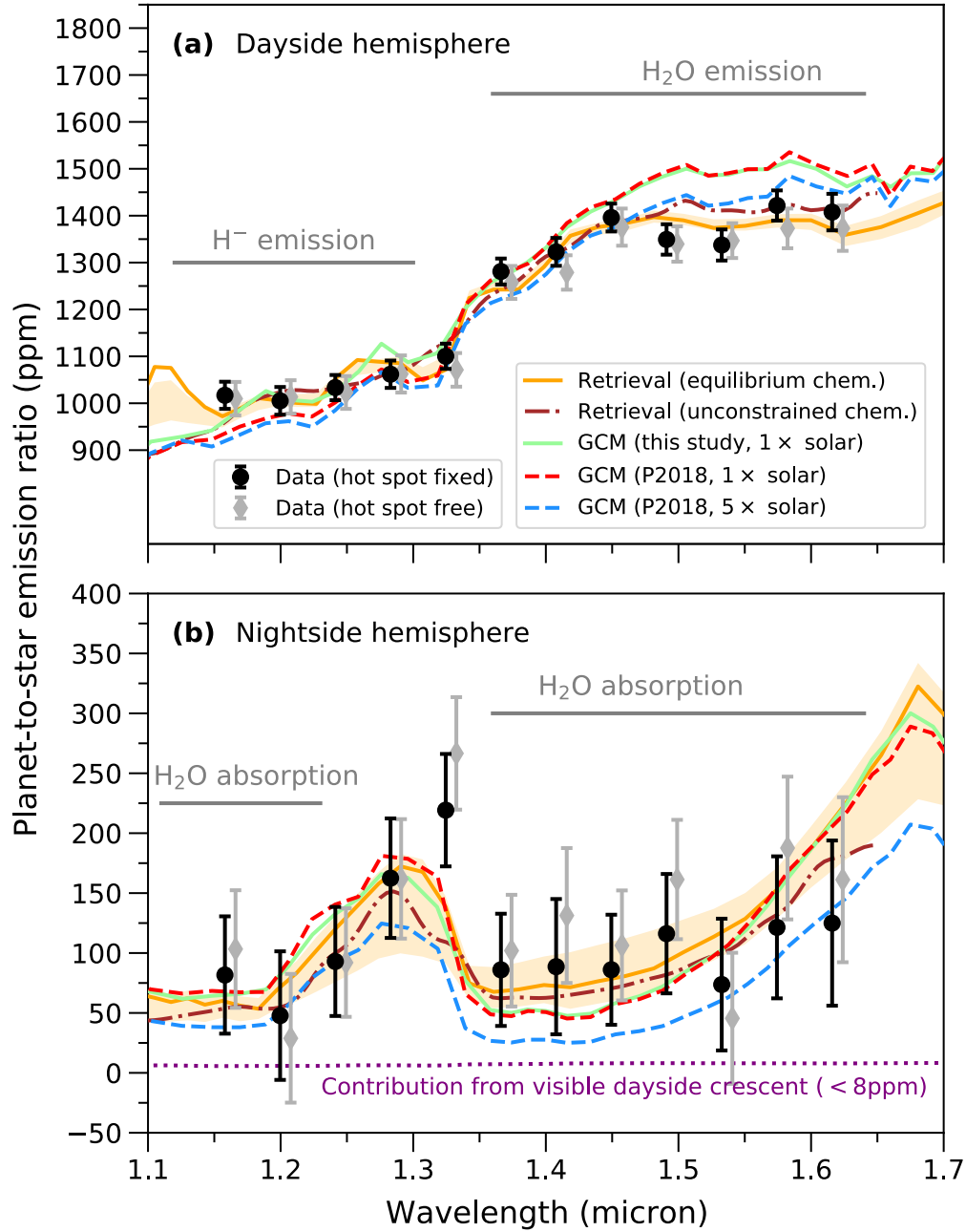

**Fig. 2. Emission spectra for the dayside and nightside hemispheres of WASP-121b. a,** Black circles show measured dayside emission with error bars corresponding to 1σ measurement uncertainties. Grey diamonds show the same, but for a light curve analysis in which the phase of maximum brightness ('hot spot') was allowed to vary in each spectroscopic channel (see Methods) and with small horizontal offsets applied for visual clarity. Orange solid line shows the maximum likelihood model and yellow shading shows the 1σ credible range of model predictions from the ATMO retrieval analysis assuming chemical equilibrium. Brown dot-dashed line shows the maximum likelihood model from the NEMESIS retrieval analysis with unconstrained chemistry (see Methods). Light green solid line shows the prediction of the 3D GCM run for this study. Dashed red and blue lines show predictions of the 3D GCMs for WASP-121b published in ref. 9 (P2018) assuming metallicities of 1× and 5× solar, respectively. **b,** The same as **a**, but showing results for the nightside hemisphere emission obtained at orbital phase 0.95, immediately prior to primary transit ingress. Dotted purple line also shows the emission contribution from the narrow crescent of dayside hemisphere visible at this orbital phase, which does not exceed 8ppm across the wavelengths covered by the data.

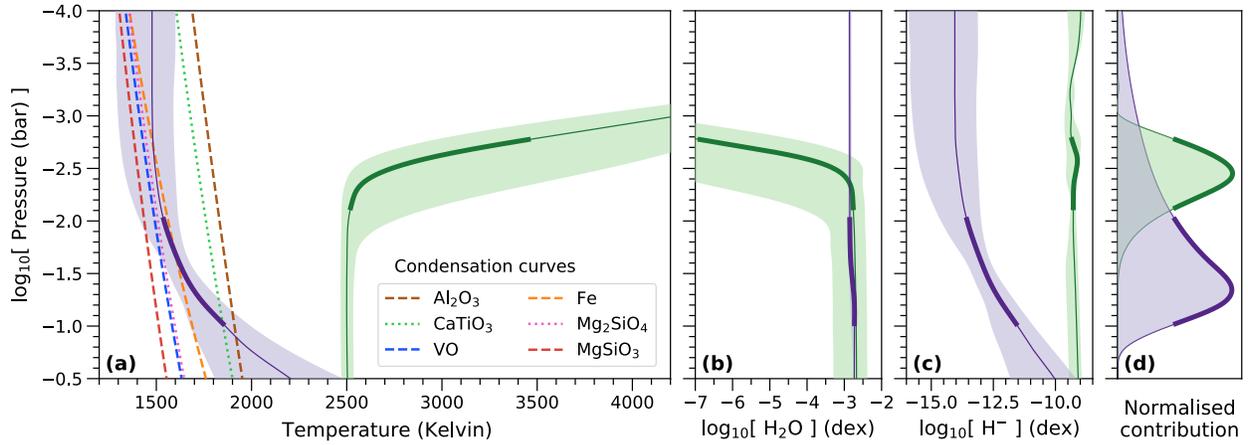

**Fig. 3. Pressure-dependent atmospheric properties retrieved for WASP-121b. a,** Green solid line shows the median temperature at each pressure level inferred for the dayside hemisphere by the ATMO retrieval analysis, with green shading showing the 1σ credible range. Purple solid line and purple shading show the same, but for the nightside hemisphere. Thick sections of the purple and green lines indicate the pressure levels from which the majority of the planetary emission is emanating at the wavelengths covered by the data. **b, c,** Vertical abundances for $H_2O$ and $H^-$, using the same colour scheme as panel **a**. The narrow range of allowed $H_2O$ abundances for the nightside is a result of the fixed metallicity. For $H^-$, the nightside abundance is more uncertain than for the dayside, because no $H^-$ spectral features are detected in the nightside spectrum. However, the $H^-$ abundance can still vary under the assumption of chemical equilibrium within the range of allowed temperatures, as the latter mediates the abundance of free electrons through thermal ionisation. **d,** Normalised contribution functions for the dayside and nightside hemispheres, integrated across the wavelength range covered by the data, using the same colour scheme as panel **a**.

## METHODS

**Observations and data reduction**

Two full orbit phase curves of WASP-121b were observed on 2018 March 12-13 and 2019 February 3-4 using the Hubble Space Telescope (HST) Wide Field Camera 3 (WFC3) infrared spectrograph with the G141 grism, covering a wavelength range of approximately 1.12-1.64μm micron. Each observation was performed over 26 consecutive HST orbits and lasted approximately 40.3 hr. The timing of each observation was designed to encompass two consecutive secondary eclipses, as these correspond to times when only the star is visible, allowing the baseline flux level to be calibrated at the beginning and end of the observation. Furthermore, both observations were scheduled such that the combined dataset provided maximum phase coverage for the planetary orbit, which has a period of 30.6 hr. Due to the long duration of the observations, full guide star reacquisition sequences were performed at the beginning of the 10th and 20th HST orbits. Integration times per exposure were 103 sec over 15 non-destructive reads (NSAMP=15) using the SPARS10 sampling sequence. Science exposures were made using the spatial scanning mode with scans along approximately 60 pixel rows of the detector cross-dispersion axis at a rate of 0.073 arcsec sec$^{-1}$. With this setup, 415 exposures were acquired for each phase curve observation. Peak frame counts were kept below ~40,000 electrons pixel$^{-1}$, within the recommended range for the detector.[36]

Spectra were extracted from each data frame using a custom Python code.[3,37-39] In brief, this involved first estimating the background flux for each exposure by taking the median count within a 10×170 pixel box away from the target on the detector. The background was then subtracted from each exposure and the target flux summed along the cross-dispersion axis within a rectangular aperture spanning 100 pixel rows, giving the target flux as a function of location along the detector dispersion axis. The mapping from the dispersion axis to wavelength was determined by cross-correlating these measured fluxes against a model stellar spectrum modulated by the G141 grism throughput.

**Broadband light curve analysis**

Broadband light curves were produced by integrating the time series of target spectra across the dispersion axis and are shown in Extended Data Fig. 1. The primary transits, secondary eclipses, and phase curve variations for the planet are easily visible by eye. However, the data are also affected by instrumental systematics caused by charge-trapping on the detector, producing a ramp-like trend in each HST orbit. In addition, longer-term instrumental drifts are evident, particularly at the beginning of the observations, and subtle discontinuities affect the measured flux at the HST orbits immediately following guide star reacquisitions.

We modelled both light curves jointly to extract properties of the planet. Our overall model $M$ took the form:

$$M_{kj} = B_{kj} \cdot \Psi_k \cdot \Pi \qquad (1)$$

where $B$ is the instrumental baseline trend, $\Psi$ is the detector ramp systematic, and $\Pi$ is the astrophysical signal. Here, the '$k$' subscripts refer to the observation number ($k$=1 for the 2018 observation and $k$=2 for the 2019 observation) and the '$j$' subscripts refer to the data segment number. For the latter, we divided each observation into three data segments, defined by the guide star re-acquisitions (Extended Data Fig. 1). In summary, we modelled: three data segments with independent baseline trends, for each observation; ramp systematics separately for each observation but shared across data segments; and a planet signal shared across all data segments of both observations.

For $B_{kj}$, we followed standard practice[7,35] and adopted a quadratic trend in time $t$ for the first data segment ($j=1$) of each visit. For the second ($j=2$) and third ($j=3$) data segments, we adopted linear $t$ trends. This was done because the baseline drift is clearly less pronounced following the first data segment (Extended Data Fig. 1), as the spacecraft and instrument has settled into a stable configuration. We also tested quadratic $t$ trends for $j=2$ and $j=3$, but found this did not improve the quality of the fit, justifying the use of the simpler linear $t$ trends.

For $\Psi_k$, we adopted the analytic treatment of ref. 40, which is motivated by a simple model of electron charge-trapping on the detector. Explicitly:

$$\Psi(t,\tau) = r \, \rho, \qquad (2)$$

where:

$$r = 1 + a_1 \exp[ -t / a_2 ], \qquad (3)$$
$$\rho = 1 + a_3 \exp[ -( t - a_4 )/( a_5 \, r ) ]. \qquad (4)$$

The astrophysical signal $\Pi$ is comprised of the combined flux received from the star-planet system:

$$\Pi = F_s + F_p, \qquad (5)$$

where $F_s$ is the emission from the host star and $F_p$ is the emission from the planet. The stellar flux $F_s$ is assumed to be constant except when the planet transits in front of the host star. We used a publicly available software package[41] to compute the drop in $F_s(t)$ during primary transit. We modelled the planet emission signal as:

$$F_p = [ \Phi + \Gamma ] E, \qquad (6)$$

where $\Phi$ denotes the phase variations, $\Gamma$ is the drop in flux received from the planet during secondary eclipse, and $E$ are ellipsoidal variations caused by tidal distortion of the planet. For $\Phi$, we used a second-order cosine function:

$$\Phi = c_0 + (c_1/2) [ 1 - \cos( \phi - c_2 ) ] + (c_3/2) [ 1 - \cos( 2\phi - c_4 ) ], \qquad (7)$$

where $\phi = 2\pi \cdot (t - T_p)/P$ is the planetary orbital phase, $P$ is the planetary orbital period, and $T_p$ is the time of mid-transit. To compute $\Gamma$, we used the same publicly available software used for the primary transit signal.[41] For $E$, we used a cosine function of the form:

$$E = 1 + (\varepsilon_0/2) [ 1 - \cos( 2\phi ) ], \qquad (8)$$

giving maximum cross-sectional area at orbital quadrature.

Using the above model, we defined a log-likelihood function for the joint dataset of the form:

$$\log P = \log N( \mathbf{y}_1 - \mathbf{M}_1, K_1 ) + \log N( \mathbf{y}_2 - \mathbf{M}_2, K_2 ) \qquad (9)$$

where $N( \boldsymbol{\mu}, K )$ denotes a multivariate normal distribution with mean vector $\boldsymbol{\mu}$ and covariance matrix $K$. In Eq. 9, the mean vector is given by the $t$-dependent model residuals, where $\mathbf{y}_k$ is a vector containing the data points and $\mathbf{M}_k$ is a vector containing the corresponding model values for the $k$th observation. The covariance matrices for each observation are assumed to be diagonal, with the form $K_{kk} = \beta_k \sigma_k I$, where $\sigma_k$ is the photon noise value, $\beta_k$ is a white noise rescaling factor, and $I$ is the identity matrix.

For the systematics components, our free parameters were the coefficients for the $t$-dependent baseline trends ($B_k$); the $a_{1,k}$, $a_{2,k}$, $a_{3,k}$, $a_{4,k}$, and $a_{5,k}$ parameters for the detector ramps ($\Psi_k$); and the white noise rescaling factors $\beta_k$. For the primary transit (i.e. $F_s$), the following parameters were allowed to vary: planet-to-star radius ratio ($R_p/R_\star$); normalised semimajor axis ($a/R_\star$); orbital impact parameter ($b = a \cdot \cos i / R_\star$, where $i$ is the orbital inclination); and the primary transit mid-times ($T_p$). These parameters were shared across both datasets, except for $T_p$ which was allowed to vary separately for each dataset. Since the planetary orbital period has been previously determined to a high level of precision, it was fixed to $P$=1.2749247646 day.[17] A circular orbit was assumed, given constraints from previous measurements.[8] A quadratic stellar limb darkening profile was adopted with coefficients ($u_1$, $u_2$) fixed to values determined using a model of the host star atmosphere, as described previously. For the phase variations $\Phi$, the parameters $c_0$, $c_1$, $c_2$, $c_3$, and $c_4$ were allowed to vary. For $\Gamma$, the secondary eclipse mid-time $T_s$ was linked to the primary transit mid-time $T_p$ according to $T_s=T_p+P/2$, given the assumed circular orbit. Rather than treating the eclipse depth as a separate free parameter, it was constrained such that $F_p$=0 at the bottom of eclipse. The ellipsoidal variation in the planetary cross-sectional area, $\varepsilon_0$, was also treated as a free parameter. Uniform priors were adopted for all free parameters. Marginalisation of the posterior distribution was performed using affine-invariant Markov chain Monte Carlo (MCMC) with 300 walkers and 1,600 steps, as implemented by a publicly available software package.[42] The best-fit model is shown in Fig. 1a and the results for the astrophysical model parameters are summarised in Extended Data Fig. 2.

In addition, a simpler first-order sinusoidal model was tested, equivalent to fixing $c_3 = c_4 = 0$ in Eq. 7. The results of this fit are also reported in Extended Data Fig. 2, with good agreement obtained for parameters common to both models. However, the first-order sinusoidal model has a significantly higher Bayesian Information Criterion (BIC) value of 2129.5 compared to 2111.0 for the second-order sinusoidal model, corresponding to a Bayes factor of $\exp(-\Delta BIC/2)=10^4$ in favour of the second-order sinusoidal model. This provides strong evidence for asymmetry in the phase curve about the 'hot spot' (i.e. phase of peak emission), which is better accounted for by the second-order sinusoidal model than by the first-order sinusoidal model. For this reason, the second-order sinusoidal model is adopted as the preferred fit for the broadband light curve.

Extended Data Fig. 2 reports the heat redistribution factors $A_F$ obtained for the first-order and second-order sinusoidal model fits. These were derived by computing $A_F=(F_{max}-F_{min})/F_{max}$ for each phase curve model sampled, where $F_{max}$ and $F_{min}$ are the maximum and minimum values of the phase curve. The obtained value for $A_F$ (95.1±2.5% for the preferred second-order sinusoid and 98.6±1.9% for the first-order sinusoid) is broadly in line with those reported for the three other gas giant exoplanets with published WFC3 phase curve measurements: $A_F$=100.5±1.3% for WASP-43b,[1,43] $A_F$=91±2% for WASP-103b,[7] and $A_F$>96% for WASP-18b.[35]

A spherical harmonic of degree $l$=2 was also used to generate a temperature map of the planet. The temperature map was converted to a corresponding phase curve signal in the G141 passband using a publicly available code.[44] Coefficients were adjusted to optimise the match of the resulting phase curve with the second-order sinusoidal function derived from the light curve fits (Supplementary Fig. 1). Since the available phase curve data do not constrain latitudinal temperature variations, coefficients of order $m$=0 were fixed to zero. The resulting temperature map is shown in Fig. 1b. At longitudes approximately 9° eastward of the substellar point, the atmosphere reaches its highest temperatures of around 3,200 Kelvin. On the nightside hemisphere, the coldest regions of the atmosphere are around 1,200 Kelvin, cool enough for numerous refractory species to condense (Fig. 3a).

**Spectroscopic light curve analysis**

Spectroscopic light curves were generated by binning the spectra into twelve wavelength channels. Before doing this, systematics common to all wavelengths were corrected using a cross-correlation technique[3,37-39] based on an original implementation by ref. 45. This common-mode correction effectively cleaned the detector ramp systematics in all but the first HST orbit of both datasets. It also substantially reduced the baseline trend systematics due to instrumental drift.

Since the common-mode correction successfully removed most of the systematics affecting each spectroscopic channel, a simpler model than described above for the broadband analysis was adopted for the light curve fitting. Rather than fit a quadratic time trend for the instrumental baseline of the first data segments (i.e. the $B_{k1}$ terms of Eq. 1), linear time trends were used for all data segments. Furthermore, with the exception of the first HST orbit of each dataset, a model for the detector ramp was unnecessary. We therefore opted to discard the first HST orbit from each spectroscopic light curve and effectively set $\Psi_k=1$ (Eq. 1) for the remaining orbits. The astrophysical signal was modelled using the same model $\Pi$ as was used for the broadband light curve fit. However, for the spectroscopic light curve fits, a number of parameters were held fixed to the best-fit values determined from the broadband light curve (Extended Data Fig. 2), namely: $T_p$, $a/R_s$, and $b$, which do not vary with wavelength, and the ellipsoidal variation amplitude, $\varepsilon_0$. Quadratic limb darkening coefficients ($u_1$, $u_2$) were determined for each spectroscopic channel as for the broadband light curve, and also held fixed during fitting. Aside from these details, light curve fitting proceeded as for the broadband light curve. The results of these fits are reported in Supplementary Table 1 and the inferred wavelength-dependent hot spot phases are shown in Supplementary Fig. 2. A second suite of spectroscopic light curve fits was also performed with the hot spot phase ($c_2$) and higher-order phase curve terms ($c_3$, $c_4$) held fixed to the maximum likelihood values derived from the broadband light curve fit. The results of these fits are reported in Supplementary Table 2. Derived heat redistribution factors $A_F$ (see above) for both suites of spectroscopic light curve fits are reported in Supplementary Table 3.

In all spectroscopic wavelength channels, the fits for which $c_2$, $c_3$, and $c_4$ were held fixed have lower BIC values than the fits for which these parameters were allowed to vary freely (Supplementary Table 3). For this reason, to derive the phase-resolved emission spectra of WASP-121b we adopt the light curve fits for which $c_2$, $c_3$, and $c_4$ were held fixed and show the corresponding best-fit light curves in Extended Data Fig. 3. To extract the phase-resolved emission spectra, the measured planetary emission for each spectroscopic light curve shown in Extended Data Fig. 3 was binned into sixteen orbital phase bins, centred at phases $\phi = 0.05$, 0.12, 0.17, 0.23, 0.28, 0.32, 0.38, 0.43, 0.57, 0.62, 0.68, 0.72, 0.78, 0.82, 0.88, and 0.95, where $\phi = 0$ coincides with the primary transit mid-point and $\phi = 0.5$ coincides with the secondary eclipse mid-point. Each phase bin had an effective width of 1.5 hours with the exception of the bins centred at $\phi = 0.05$ and $\phi = 0.95$, which had larger widths of 3 hours to compensate for the lower fluxes at those phases. Uncertainties were calculated as the standard deviation of model residuals within each phase bin added in quadrature to the standard deviation of in-eclipse model residuals (i.e. the uncertainty in the stellar baseline flux level), following ref. 7. The planetary emission measurements obtained in this way are reported in Supplementary Table 4 and plotted in Extended Data Figure 6. As described in the main text, a dayside emission spectrum was also generated from the distribution of light curve emission maxima generated during the fitting. This dayside spectrum is reported in Supplementary Table 5 and, as shown in Supplementary Fig. 3, is in excellent agreement with the spectrum obtained from the emission measured in the phase bin immediately preceding secondary eclipse ingress. It also has a similar shape but a slightly higher overall level than the emission measured in the phase bin immediately following eclipse egress. Both of these observations are consistent with expectations, as the phase curve peak coincides with the phase bin immediately preceding eclipse ingress (Fig. 1 and Extended Data Fig. 2). Furthermore, the secondary eclipse spectrum presented by ref. 3 also has a similar shape, but intermediate overall level, relative to these spectra (Supplementary Fig. 3). Again, this is to be expected, as the secondary eclipse depths of ref. 3 were measured relative to an out-of-eclipse baseline that was linear in $t$ and thus effectively the average of the emission measured immediately before and after eclipse.

**Retrieval analyses assuming chemical equilibrium**

Atmospheric retrieval analyses were performed on the phase-resolved emission spectra using ATMO, a one-dimensional radiative transfer code used to simulate substellar atmospheres.[46-54] ATMO solves the radiative transfer equation in plane-parallel geometry assuming hydrostatic and radiative-convective equilibrium. The first step of the analysis was to perform a retrieval on the dayside spectrum derived from the measured emission maxima in each spectroscopic wavelength channel (Supplementary Table 5). For this retrieval, the pressure-temperature (PT) profile was freely fit using the analytic profile of ref. 55, with three free parameters: the infrared opacity ($\kappa_{IR}$); the ratio of the visible-to-infrared opacity ($\gamma=\kappa_V/\kappa_{IR}$); and an irradiation efficiency factor ($\psi$). The atmosphere was assumed to be in chemical equilibrium, with the heavy element abundances (i.e. metallicity) varied as a free parameter ([M/H]). Chemical abundances were calculated using Gibbs energy minimisation for 175 gaseous species, 9 ionic species, and 93 condensate species.[51-54] Rainout of condensed species consistent with the retrieved PT profiles was included,[51,52] as was thermal ionisation and dissociation. Opacities for the spectrally active species $H_2O$, $CO_2$, CO, $CH_4$, $NH_3$, Na, K, Li, Rb, Cs, TiO, VO, FeH, $PH_3$, $H_2S$, HCN, $C_2H_2$, $SO_2$, Fe, and $H^-$ were included, along with collision-induced absorption due to $H_2$–$H_2$ and $H_2$–He. Uniform priors were adopted for all model parameters ($\kappa_{IR}$, $\gamma$, $\psi$, [M/H]) and fitting was performed using nested sampling.[56-58] A PHOENIX BT-Settl model[59] was adopted for the stellar spectrum, assuming a stellar effective temperature $T_\star$=6,500 Kelvin, surface gravity log $g_\star$=4.0 cm·s$^{-2}$, and radius $R_\star$=1.458 solar radii, based on the values provided by ref. 8. The resulting posterior distributions are reported in Extended Data Fig. 4 and the corresponding emission spectrum distribution is shown in Fig. 2a. The maximum likelihood model has a $\chi^2$ value of 9.74 for 8 degrees of freedom (i.e. reduced $\chi_v^2 = 1.2$), indicating a good fit to the data. Also shown in Fig. 3 are posterior distributions for the PT profile (Fig. 3a), $H_2O$ abundance (Fig. 3b), and $H^-$ abundance (Fig. 3c), and the contribution function for the maximum likelihood model (Fig. 3d). Note that the $H_2O$ and $H^-$ abundances were determined from the chemical equilibrium abundances and were not fit directly as free parameters.

For the remaining phase-resolved emission spectra, varying fractions of the dayside and nightside hemispheres are visible. Due to the strong contrast in effective temperature between each hemisphere, the dayside and nightside spectra are expected to differ substantially. To accommodate this, retrievals were performed using a method similar to the "2TP-Fixed" framework described by ref. 22. Under this approach, denoted here as "2x PT", the combined emission received from the planet $\Phi$ at each orbital phase was assumed to be described by:

$$\Phi = \eta_d \Phi_d + (1 - \eta_d)\Phi_n \qquad (10)$$

where $\eta_d$ is the fractional area of the visible dayside hemisphere, $\Phi_d$ is the planetary emission from the dayside hemisphere, and $\Phi_n$ is the planetary emission from the nightside hemisphere. The fractional area of the visible dayside hemisphere $\eta_d$ is given by:

$$\eta_d = [1 - \cos(2\pi\phi + \pi - c_2)]/2 = [1 + \cos(2\pi\phi - c_2)]/2 \qquad (11)$$

This is a slight variation of Equation A2 in ref. 22, with $c_2$ corresponding to the phase of maximum brightness and set to the value obtained from the broadband light curve fit (Extended Data Fig. 2). Given that the data are sensitive to thermal emission, rather than reflected light, the inclusion of the $c_2$ offset accounts for the overall advection of gas prior to re-emission. The dayside emission $\Phi_d$ was also held fixed to the maximum likelihood model described above and shown in Fig. 2a. This was done because the dayside spectrum derived from the emission maxima has a relatively high signal-to-noise and retrieving for both the dayside and nightside contributions at each orbital phase was not justified given the limited number of data points (i.e. twelve spectroscopic channels). The metallicity was also assumed to be the same for both hemispheres and held fixed to [M/H]=0.7 (i.e. 5× solar), close to the median value derived from the dayside spectrum (Extended Data Fig. 4). With the metallicity and dayside emission held fixed, this left the three PT profile parameters ($\kappa_{IR}$, $\gamma$, $\psi$) for the nightside hemisphere as the remaining free parameters. As for the initial retrieval for the dayside spectrum described above, fitting was again performed using nested sampling, with uniform priors for the PT profile parameters. The nightside PT profiles were retrieved in this way for phases $\phi$ = 0.05, 0.12, 0.17, 0.23, 0.28, 0.72, 0.78, 0.82, 0.88, and 0.95. Useful constraints could not be obtained by fitting for the nightside PT profiles at the remaining phases (i.e. $\phi$ = 0.32, 0.38, 0.43, 0.57, 0.62, 0.68), as the nightside emission $\Phi_n$ comprised a relatively small fraction of the total planetary emission. For these latter phases, $\Phi_n$ was instead held fixed to the maximum likelihood model obtained for the $\phi$ = 0.95 retrieval (Fig. 2b), as the nightside emission had the highest signal-to-noise at this phase, and the PT profile was instead allowed to vary for the dayside emission component $\Phi_d$. Resulting phase-resolved emission spectra are shown for all phases in Extended Data Fig. 6. The corresponding PT profiles, $H_2O$ abundances, and $H^-$ abundances are shown in Extended Data Figs 7, 8, and 9, respectively. The $\chi^2$ fit statistics are reported in Supplementary Table 6, with a mean reduced $\chi_\nu^2 = 1.1$ and median reduced $\chi_\nu^2 = 1.2$ achieved across the sixteen phase bins. One final retrieval was performed for phase $\phi$ = 0.95, the same as before (i.e. dayside contribution held fixed and PT parameters allowed to vary) but with the metallicity also allowed to vary as a free parameter. The results of this retrieval are given in Extended Data Fig. 4 and the inferred metallicity ([M/H]=$0.66^{+0.70}_{-1.02}$) is found to be consistent with the [M/H]=0.7 (i.e. 5× solar) value assumed for the fiducial retrievals, providing a useful validation of the latter.

To assess the significance of the nightside emission detections, the $\chi^2$ and BIC values were computed under the assumption of zero nightside emission (i.e. $\Phi_n = 0$) at the ten phases for which the "2x PT" retrievals were performed for the nightside hemisphere. As reported in Supplementary Table 7, the BIC values of the "2x PT" retrievals are lower than those of the retrievals assuming $\Phi_n = 0$ for seven of the ten phases considered. At three of the individual phases ($\phi$ = 0.23, 0.88, 0.95) the Bayes factors are >100, corresponding to decisive preference for the "2x PT" models over the $\Phi_n = 0$ null hypothesis.[60,61] The strongest preference for the "2x PT" retrieval is seen at phase $\phi$ = 0.95, for which the null hypothesis is disfavoured by a Bayes factor of 7.1 × $10^{10}$. This translates to a preference for the "2x PT" model over the $\Phi_n = 0$ null hypothesis at a significance in excess of 5σ under the frequentist paradigm, following the conversion provided by ref. 62. If all ten phases are considered together as an ensemble, the BIC is 221.3 for the "2x PT" retrievals and 380.1 for the $\Phi_n = 0$ null hypothesis (last row of Supplementary Table 7), amounting to an overall rejection of the null hypothesis with a Bayes factor of 3×$10^{34}$. The unambiguous preference for the "2x PT" models over the $\Phi_n = 0$ null hypothesis at multiple phases implies that emission from the nightside hemisphere is distinct from the dayside contribution and detected at high confidence in the data.

As an additional check, simple blackbody spectra were fit to each phase-resolved emission spectrum. For these fits, the effective planetary temperature was the only free parameter. The maximum likelihood spectra are plotted in Extended Data Fig. 6 and the results are reported in

Supplementary Table 6. These blackbody fits had a mean reduced $\chi_\nu^2 = 2.1$ and median reduced $\chi_\nu^2 = 2.0$ across the sixteen phase bins, which is significantly poorer than those obtained for the "2x PT" fits (Supplementary Table 6). However, the brightness temperatures derived from the blackbody fits for the dayside and nightside hemispheres allow simple estimates to be made for the planetary Bond albedo ($A_B$) and heat redistribution efficiency ($\varepsilon$) following the method of ref. 63. The results of this analysis are shown in Supplementary Fig. 4a. As can be seen in Supplementary Fig. 4b, the Bond albedo derived for WASP-121b ($A_B = 0.14 \pm 0.08$) is consistent with values reported for other hot Jupiters that have similar irradiation temperatures ($T_0 = T_\star\sqrt{R_\star/a} = 3320 \pm 72$ K for WASP-121b). However, the derived value for the heat redistribution efficiency ($\varepsilon = 0.29 \pm 0.02$ for WASP-121b) is notably higher than for two of those shown in Supplementary Fig. 4c with similar irradiation temperatures: namely, WASP-18b ($T_0 = 3412 \pm 49$, $\varepsilon = 0.01^{+0.07}_{-0.01}$ )[64] and KELT-1b ($T_0 = 3391 \pm 29$, $\varepsilon = 0.06^{+0.03}_{-0.02}$ ).[64] These measurements hint at a diversity of circulation regimes among highly-irradiated substellar objects.

**Retrieval analyses with unconstrained chemistry**

A second suite of retrieval analyses was performed using the NEMESIS radiative transfer and retrieval model.[23,65-68] NEMESIS couples a parametric, one-dimensional radiative transfer simulation to the PyMultiNest algorithm[69,70] which uses nested sampling to explore the model parameter space.[56-58] The correlated-k approximation[71] is used to pre-tabulate gas absorption data.

There were a number of important differences between the approaches used by ATMO and NEMESIS to model the WASP-121b atmosphere. First, the NEMESIS retrievals only fit for the abundances of $H_2O$ and $H^-$; the two other main species expected to be spectrally active at the wavelengths probed by the data are VO and FeH (Extended Data Fig. 5), which were included with a constant mole fraction fixed to the values inferred by ref. 72. Second, the abundances of $H_2O$ and $H^-$ were allowed to vary freely at each orbital phase, without the requirement of satisfying chemical equilibrium. Third, following ref. 9, a simple analytic treatment was adopted to account for thermal dissociation of $H_2O$ at phases for which the properties of the dayside atmosphere were retrieved (i.e. $\phi = 0.32, 0.38, 0.43, 0.57, 0.62, 0.68$). Specifically, the following parameterisation for the $H_2O$ mole fraction ($X$) as a function of pressure ($P$) was adopted:

$$X(P) = \begin{cases} X_{\text{deep}} & \text{for } P \geq P_{\text{knee}} \\ X_{\text{deep}}(P/P_{\text{knee}})^\alpha & \text{for } P < P_{\text{knee}} \end{cases} \quad (12)$$

where the deep atmosphere $H_2O$ mole fraction ($X_{\text{deep}}$), knee pressure ($P_{\text{knee}}$), and power law index ($\alpha$) were fitted as free parameters. As in ref. 72, $H^-$ was assumed to be well-mixed throughout the atmosphere, with constant mole fraction retrieved as an additional parameter. The remaining atmosphere was assumed to be composed of $H_2$ and He in a 9:1 ratio.

The $H_2O$, VO, and FeH k-tables were computed according to ref. 73, using the data presented in refs. 74, 75, and 76, respectively. The $H^-$ bound-free and free-free opacities were calculated according to ref. 77. Also included were $e^-$ and H, both assumed to be well-mixed with abundances fixed to the deep atmosphere $H^-$ abundance. The latter was justified because the abundances of $e^-$ and H do not affect the observed emission spectrum provided that the abundances are sufficiently high to allow vigorous interaction with the $H^-$ ions. Collision-induced absorption due to $H_2$–$H_2$ and $H_2$–He was also included.[78-82] A parameterisation identical to that used by the ATMO retrievals was adopted for the PT profile. At each orbital phase, this gave a final model with seven free parameters: three parameters for the PT profile ($\kappa_{IR}$, $\gamma$, $\psi$); the $H^-$ mole fraction;

and the three parameters defined above for the pressure-dependent $H_2O$ mole fraction ($X_{\text{deep}}$, $P_{\text{knee}}$, $\alpha$).

Results for the fiducial dayside and nightside hemisphere PT profiles, $H_2O$ abundances, and $H^-$ abundances are compared to those obtained by the ATMO retrievals in Extended Data Fig. 10. Maximum likelihood emission spectra are shown for all orbital phases in Extended Data Fig. 6, with the PT profiles, $H_2O$ abundances, and $H^-$ abundances that were retrieved separately for each phase shown in Extended Data Figs 7, 8, and 9, respectively. Overall, the PT profiles inferred by NEMESIS for the nightside phases are in good agreement with those inferred by ATMO. The agreement is reasonable, but not as good, for the dayside phases. The latter is likely due to the challenge of accounting for thermal dissociation and ionisation using the parameterised approach described above for the free chemistry NEMESIS retrievals. The NEMESIS retrievals at dayside phases do not succeed in accounting for the thermal dissociation of $H_2O$ (Extended Data Fig. 10b). Instead, to account for the muted $H_2O$ spectral band, the NEMESIS retrieval favours a higher $H^-$ abundance compared to ATMO (Extended Data Fig. 10c), which raises the opacity at wavelengths shortward of 1.3µm in particular (Extended Data Fig. 5). The overall raised opacity produces an extended wing in the contribution function towards lower pressures (Extended Data Fig. 10d), in turn favouring a thermal inversion at lower pressures than ATMO (Extended Data Fig. 10a). This also explains why NEMESIS infers thermal inversions at lower pressures than ATMO for phases $\phi = 0.43$, 0.57, and 0.62 (Extended Data Fig. 7).

However, the broad agreement between the NEMESIS and ATMO results is reassuring, given the different methodologies adopted. The NEMESIS $\chi^2$ fit statistics are reported alongside those for ATMO in Supplementary Table 6. For NEMESIS, the mean reduced $\chi^2_\nu = 2.43$ and the median reduced $\chi^2_\nu = 2.34$, which are significantly higher than the equivalent fit quality metrics achieved by ATMO. This is primarily a consequence of the larger number of parameters required for the NEMESIS retrievals (i.e. seven for NEMESIS versus three for ATMO), although the absolute $\chi^2$ values are also higher, indicating that the NEMESIS models do not replicate the data as well as the ATMO models overall. The latter is due to the failure of NEMESIS to adequately treat the thermal dissociation of $H_2O$ for the dayside spectra, as noted above. For these reasons, along with the physically-motivated enforcement of chemical equilibrium, we present the ATMO retrievals as our primary analysis (Figs 2 and 3).

**General circulation models**

A three-dimensional general circulation model (GCM) simulation was performed for the atmosphere of WASP-121b using the Substellar and Planetary Radiation and Circulation (SPARC) model.[9,83-91] The model couples the MITgcm dynamical core,[92] a finite-volume code that solves the three-dimensional primitive equations on a staggered Arakawa C grid,[93] with a plane-parallel, two-stream version of a multi-stream radiation code developed for planetary atmospheres.[94] Opacities are calculated using the correlated-$k$ method[95] assuming local thermodynamic and chemical equilibrium for each PT point, using the solar photosphere elemental abundances of ref. 96. In particular, the model includes opacity due to important absorbers such as $H_2O$, $H^-$, CO, TiO, and VO, but does not yet include atomic metals such as Fe and Mg. The coupling of the dynamical core and radiative transfer scheme allow for the self-consistent calculation of the heating and cooling rates of the atmosphere.

The SPARC GCM for WASP-121b has a horizontal resolution of C32 (128×64 in longitude and latitude, respectively) and a vertical resolution of 45 pressure levels evenly spaced in log pressure, that extend from a mean pressure of 1,000 bar at the bottom to 200 μbar at the top. The model was integrated for 80 Earth days (~60 planetary orbits). A global map of the temperature and wind speeds at a pressure of 10 mbar (a pressure within the range of altitudes probed at the WFC3 wavelengths, e.g. Fig. 3d) is shown in Supplementary Fig. 5. The map shows predominantly eastward flow at the equator and nightside vortices, with dayside temperatures exceeding 3,000 Kelvin and nightside temperatures dropping to ~1,000 Kelvin. Synthetic phase curves were generated from the GCM following refs 97 and 98, and are shown in Fig. 1 and Extended Data Fig. 3. Predicted emission spectra are shown in Fig. 2 and Extended Data Fig. 6. Also shown in the latter figures are predicted emission spectra from the independent GCM simulations of ref. 9, which were performed for atmospheric metallicities of 1× and 5× solar.

Two important caveats are worth highlighting, which apply to the GCM simulation presented here, as well as those published by ref. 9. First, the lack of opacity due to metals such as Fe and Mg could be important, as separate modelling has shown that these metals can play major roles in determining the outgoing emission for ultrahot Jupiters such as WASP-121b.[10] Second, the atmosphere of WASP-121b was assumed to be cloud-free to simplify the modelling. However, clouds could potentially play a substantial role in the atmospheric radiative transfer for WASP-121b, particularly on the nightside hemisphere and in the terminator region where temperatures are relatively low and likely conducive to the condensation of numerous species (Fig. 3b).

**Data availability**

Raw HST data frames are publicly available online at the Mikulski Archive for Space Telescopes (MAST; https://archive.stsci.edu). Light curves and the extracted emission spectra are provided as plain text files in the Source Data. Data shown in the main article figures and Extended Data figures are provided in the accompanying Source Data. Additional data products (time series spectra extracted from each exposure and the raw spectroscopic light curves) are provided as Supplementary Data.

**Code availability**

The main analysis routines have been written by the authors in Python and are available on request. Other publicly available code that was used has been cited throughout the text.

**ACKNOWLEDGEMENTS**

The authors are grateful to Vivien Parmentier for helpful discussion and providing published models for the atmosphere of WASP-121b. Support for HST program GO-15134 was provided by NASA through a grant from the Space Telescope Science Institute, which is operated by the Association of Universities for Research in Astronomy, Inc., under NASA contract NAS 5-26555. T.M.E. acknowledges financial support from the Max Planck Society. J.B. is supported by a Science and Technology Facilities Council Ernest Rutherford Fellowship. J.T. acknowledges financial support from the Canadian Space Agency. N.J.M. is supported by a UKRI Future Leaders Fellowship: MR/T040866/1, a Science and Technology Facilities Council Consolidated Grant (ST/R000395/1) and the Leverhulme Trust through a research project grant (RPG-2020-82).

## AUTHOR CONTRIBUTIONS STATEMENT

T.M.E. co-led the observing program, analysed the HST data, and wrote the manuscript. D.K.S., J.K.B, and J.T. performed the retrieval analyses. T.K. co-led the observing program and ran the GCM simulation. J.G. and N.L. processed the GCM results. T. D. assisted with the light curve analysis. All authors discussed the results and provided feedback on the manuscript.

## COMPETING INTERESTS STATEMENT

The authors declare no competing interests.

## EXTENDED DATA

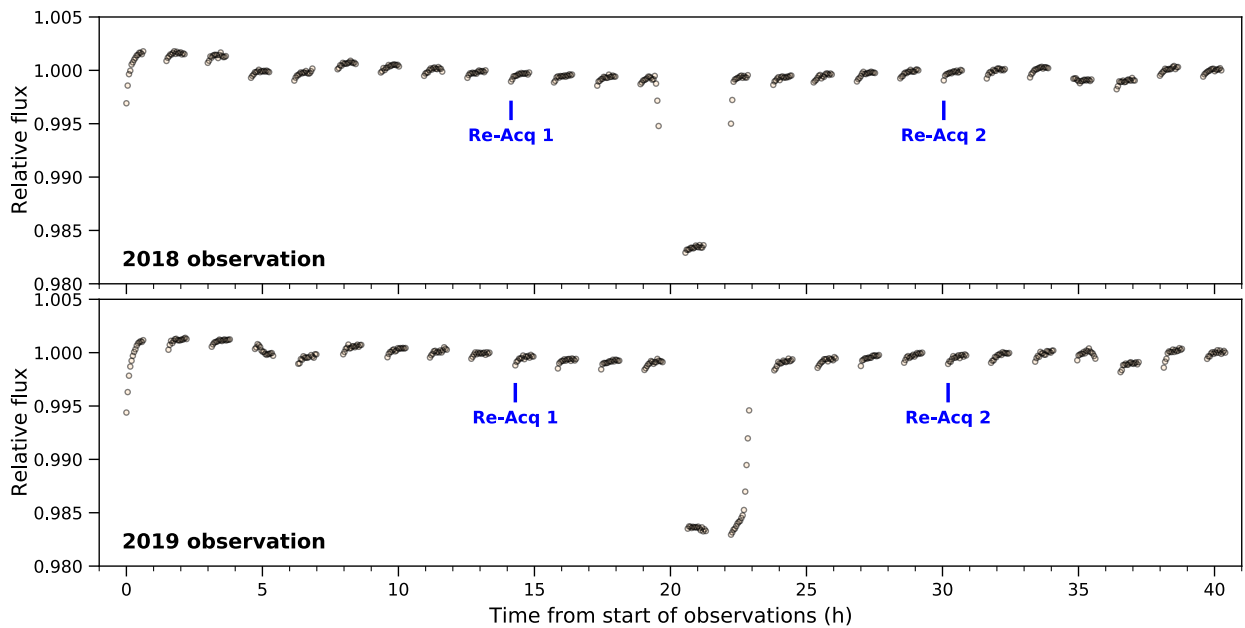

**Extended Data Fig. 1. Raw broadband light curves for WASP-121b.** WFC3 observations made in 2018 and 2019. Gaps in the time series are due to the target disappearing from view for approximately half of each HST orbit. Two secondary eclipses and the primary transit are visible by eye. The data are affected by detector systematics that result in a ramp-up of flux registered during each HST orbit. Guide star re-acquisitions were performed at the beginning of the 10th and 20th HST orbits for both observations.

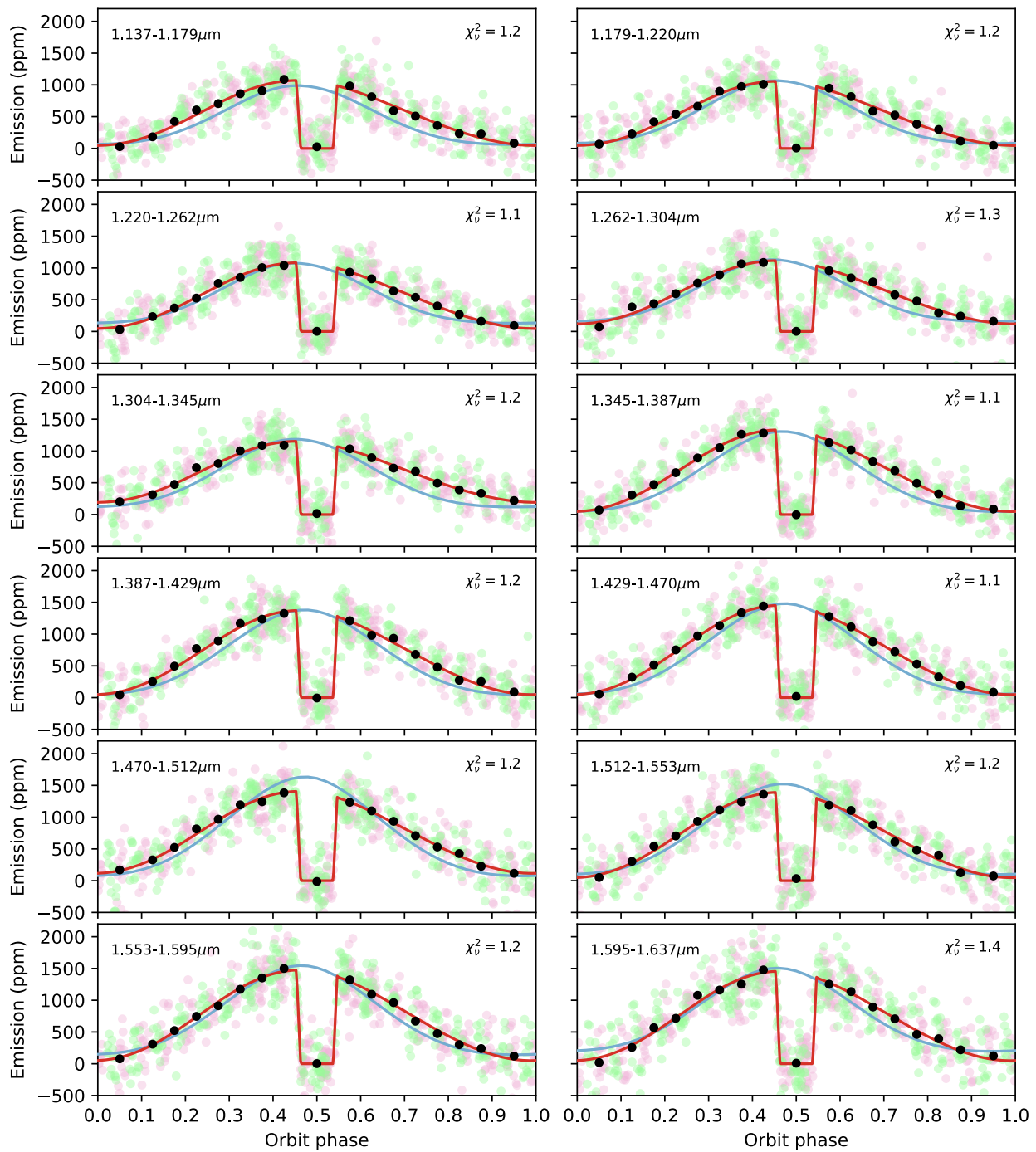

**Extended Data Fig. 2. Spectroscopic phase curves for WASP-121b.** Systematics-corrected spectroscopic phase curves measured with WFC3. Pink circles show 2018 data and green circles show 2019 data. Black circles show the combined dataset binned in phase, with marker sizes approximately corresponding to the measurement uncertainties. Phase bins are the same as those used for generating the phase-resolved emission spectra, plus an additional in-eclipse bin. Red lines show the maximum likelihood second-order sinusoidal fits, with corresponding reduced $\chi_\nu^2$ values listed in the upper right corner of each axis. Blue lines show predictions of the 3D GCM run for the present study.

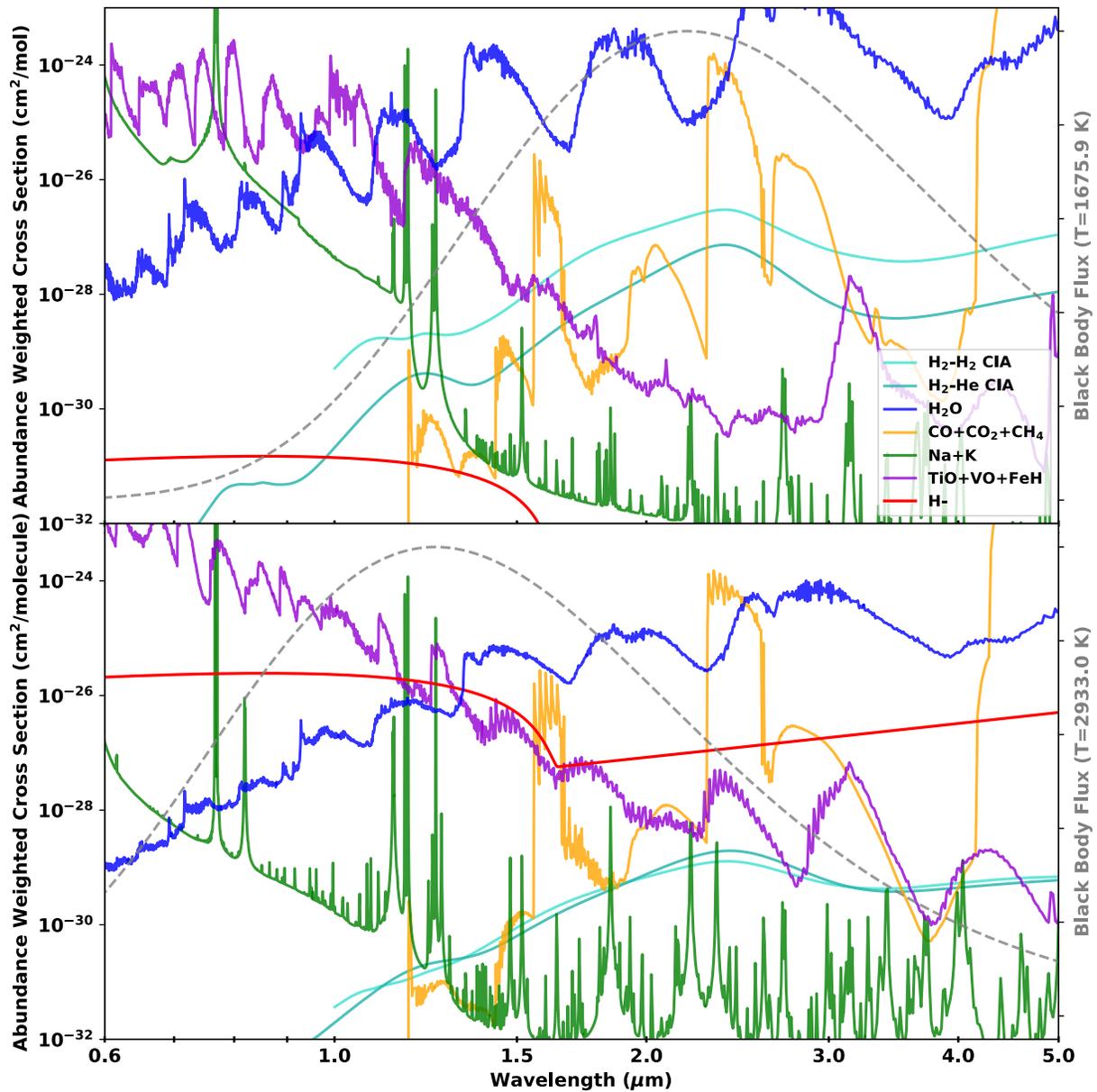

**Extended Data Fig. 3. Spectrally active gases on the nightside (top panel) and dayside (bottom panel) hemispheres.** Solid lines show absorption cross-sections for spectrally active species weighted by their mole fractions at pressures just below the contribution function peaks, as inferred by the ATMO "2x PT" retrieval analyses. Dashed grey lines show blackbody emission curves for the retrieved temperatures at the same pressure levels.

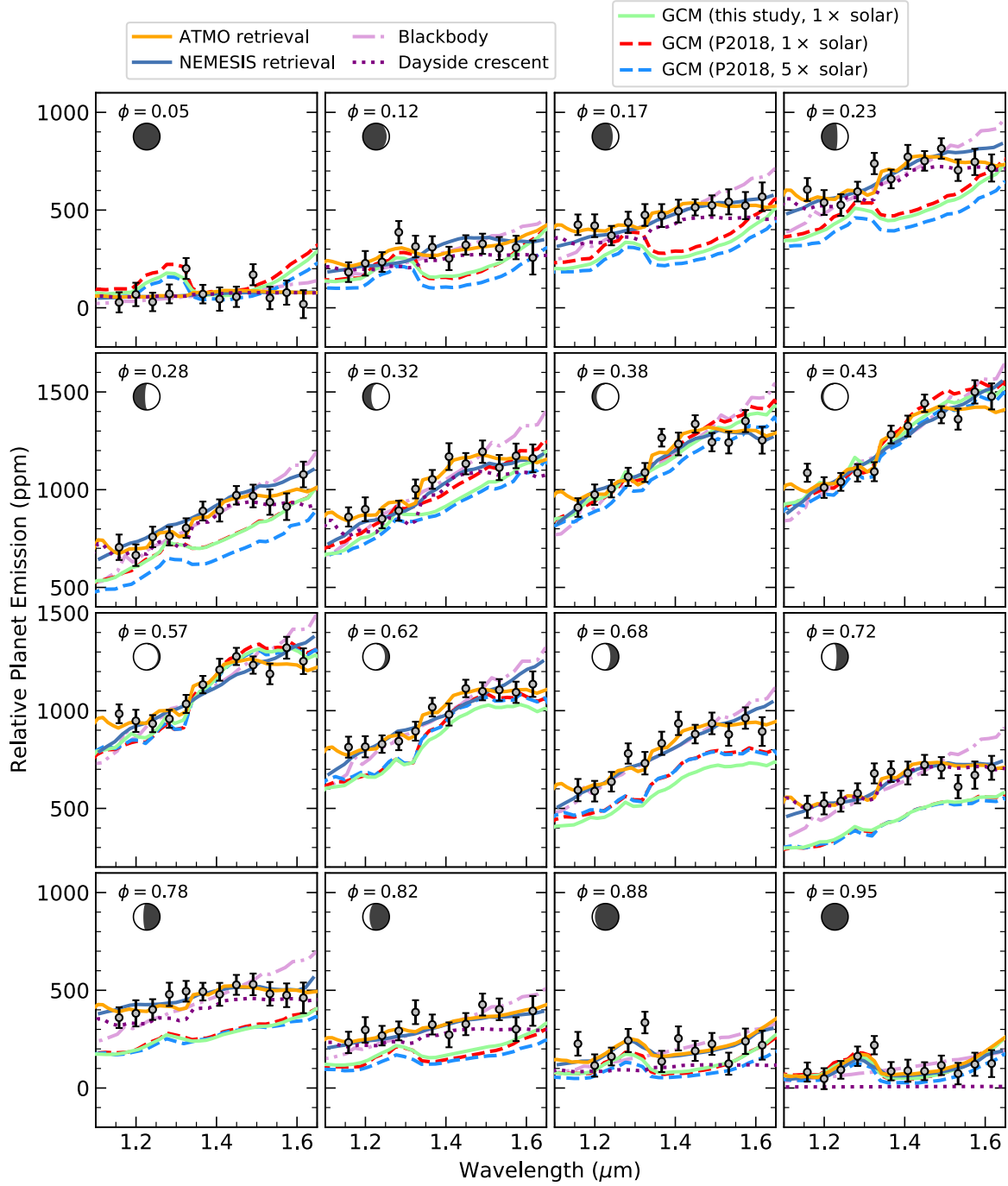

**Extended Data Fig. 4. Emission spectra at different orbital phases for WASP-121b.** Grey circles show measured planet emission as a function of wavelength with error bars indicating 1σ measurement uncertainties. Solid orange and dark blue lines show the maximum likelihood spectra obtained from the ATMO "2x PT" retrievals and NEMESIS "2x PT" retrievals, respectively. Dotted purple lines show dayside contributions for the phases at which retrievals were performed for the nightside emission. Solid light green line shows the emission predicted by the 3D GCM run for the present study assuming 1x solar metallicity. Dashed red and blue lines show the emission predicted by the 3D GCM simulations of ref 10 assuming 1x and 5x solar metallicity, respectively. Dot-dashed purple lines show best-fit blackbody spectra. Circle symbols indicate the illuminated fraction of the visible hemisphere at each orbital phase.

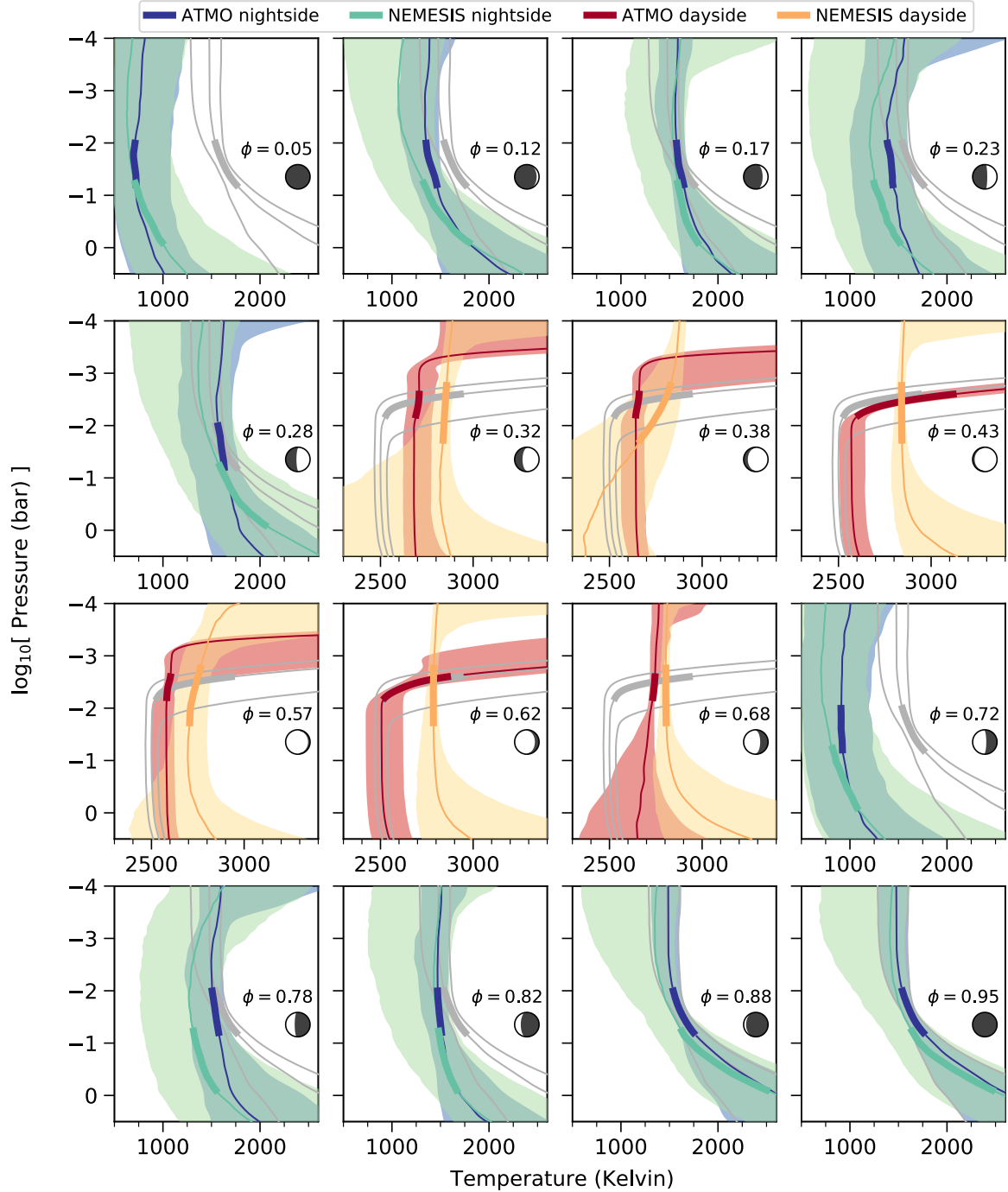

**Extended Data Fig. 5. Pressure-temperature (PT) profiles retrieved for different orbital phases**. Blue lines show the median PT profiles obtained from the ATMO "2x PT" nightside retrievals, with blue shading the corresponding 1σ credible ranges. Red lines and shading show the same for the ATMO "2x PT" dayside retrievals. Green and yellow lines and shading show the same for the NEMESIS "2x PT" nightside and dayside retrievals, respectively. For the ATMO retrievals, thick lines correspond to the same pressures highlighted in Fig. 3, where the contribution function is greatest. For the NEMESIS retrievals, thick lines indicate the equivalent pressures of greatest contribution for those retrievals (Extended Data Fig. 8d). Grey lines reproduce the fiducial dayside and nightside PT distributions of Fig. 3.

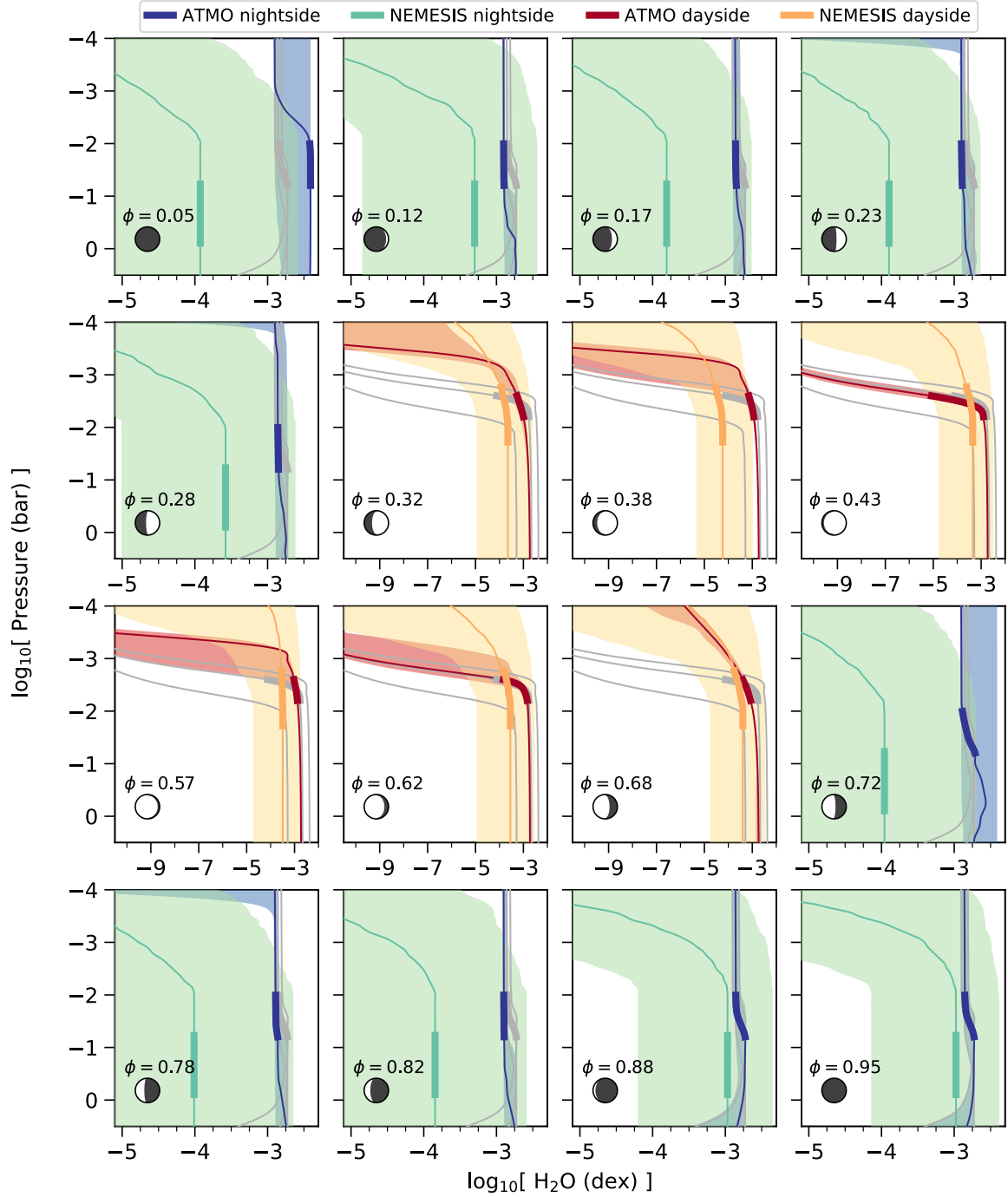

**Extended Data Fig. 6. Retrieved pressure-dependent H₂O abundances.** Same format as Extended Data Figure 5, but showing posterior distributions for the retrieved H$_2$O abundances. ATMO credible ranges are much narrower than those of NEMESIS, due to the metallicity being fixed for the ATMO retrievals whereas the H$_2$O abundance was unconstrained for the NEMESIS retrievals.

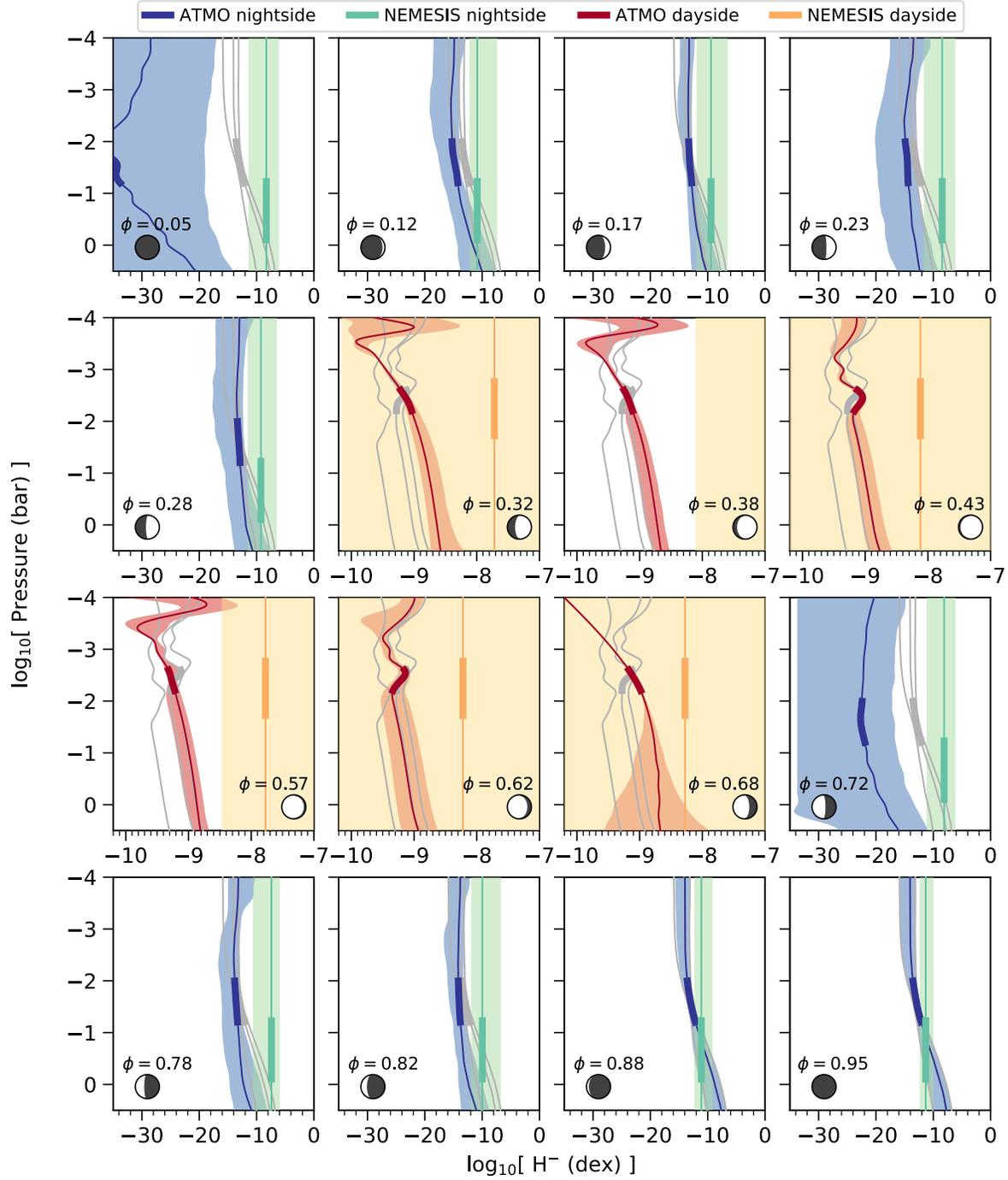

**Extended Data Fig. 7. Retrieved pressure-dependent H⁻ abundances.** Same format as Extended Data Figs 5 and 6, but showing posterior distributions for the retrieved H⁻ abundances.

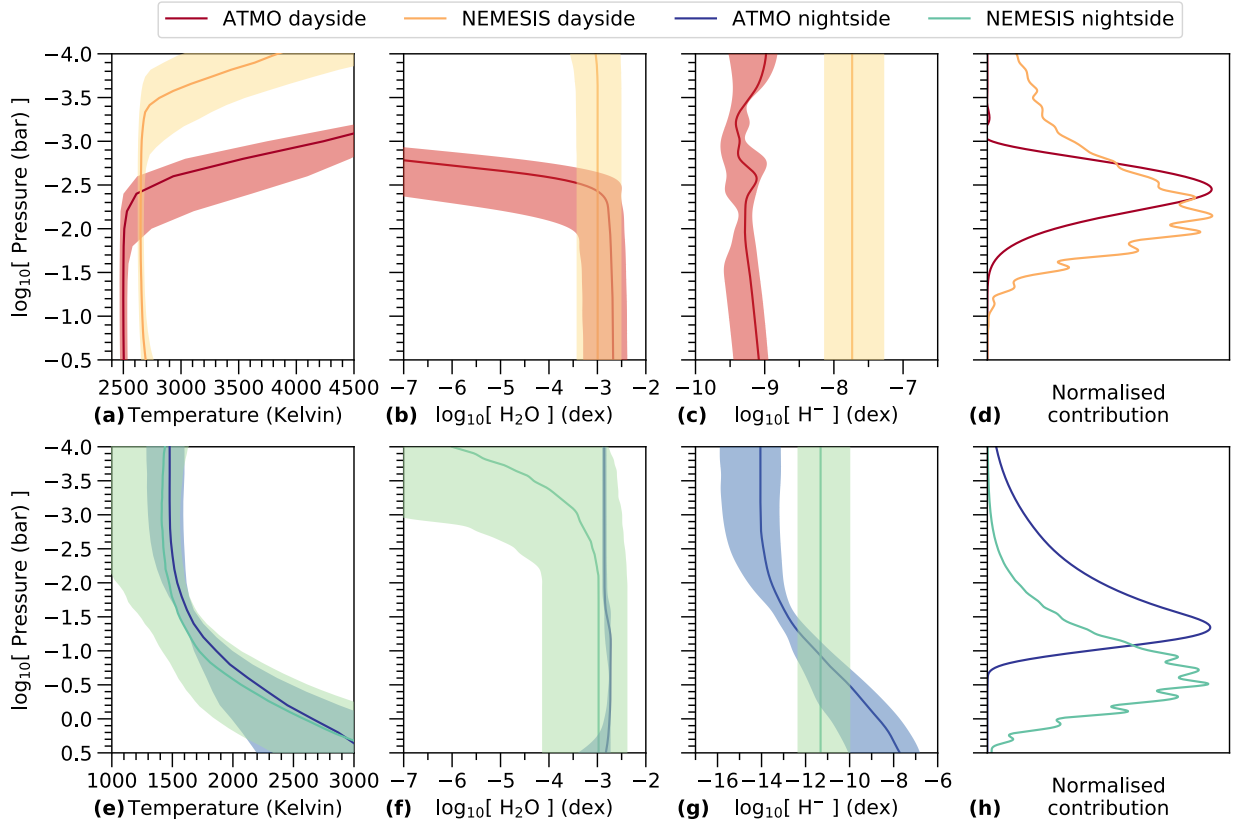

**Extended Data Fig. 8. ATMO and NEMESIS "2x PT" retrievals for the fiducial dayside and nightside spectra.** Top panel shows (from left to right) the retrieved PT profiles, H$_2$O abundances, H$^-$ abundances, and normalised contribution functions for the fiducial dayside spectrum. Bottom panel shows the same for the fiducial nightside spectrum. Same colour scheme as Extended Data Figs 5-7.

**Extended Data Table 1. Astrophysical parameters obtained from fits to the broadband light curve.** Results are reported for fits adopting first-order and second-order sinusoidal models for the planetary phase variations. Values listed for free parameters are posterior medians and uncertainties give the 1σ credible ranges. Note that inclination $i$ values have been derived from $b$ and $a/R_\star$. The redistribution factors $A_F$ have been derived from the distribution of phase curve models sampled during fitting. Although the white noise was treated as a free parameter for each dataset as described in Methods, the quoted $\chi^2$ and BIC values were calculated assuming photon noise to allow direct comparison of the two models.

| Parameter | Unit | Sinusoid first-order $\chi^2 = 1896.6$, $\chi_v^2 = 2.6$ BIC = 2129.5 | Sinusoid second-order $\chi^2 = 1864.9$ $\chi_v^2 = 2.5$ BIC = 2111.0 |
|---|---|---|---|
| $R_p/R_\star$ | – | $0.12190^{+0.00009}_{-0.00008}$ | $0.12199^{+0.00009}_{-0.00009}$ |
| $a/R_\star$ | – | $3.812^{+0.004}_{-0.003}$ | $3.785^{+0.003}_{-0.003}$ |
| $b$ | – | $0.114^{+0.002}_{-0.003}$ | $0.175^{+0.002}_{-0.002}$ |
| $i$ | degree | $88.29^{+0.05}_{-0.04}$ | $87.35^{+0.03}_{-0.02}$ |
| $T_{p,1}$ | JD$_\text{UTC}$ | $2458191.114061^{+0.000049}_{-0.000050}$ | $2458191.114045^{+0.000054}_{-0.000048}$ |
| $T_{p,2}$ | JD$_\text{UTC}$ | $2458518.768537^{+0.000073}_{-0.000065}$ | $2458518.768609^{+0.000072}_{-0.000060}$ |
| $c_0$ | ppm | $18^{+28}_{-14}$ | $26^{+34}_{-19}$ |
| $c_1$ | ppm | $1195^{+23}_{-29}$ | $1158^{+28}_{-40}$ |
| $c_2$ | radian | $3.06^{+0.04}_{-0.05}$ | $3.03^{+0.03}_{-0.04}$ |
| $c_3$ | ppm | – | $69^{+23}_{-23}$ |
| $c_4$ | radian | – | $4.54^{+0.03}_{-0.02}$ |
| $\varepsilon_0$ | percent | $1.1^{+1.6}_{-0.8}$ | $1.6^{+2.3}_{-1.2}$ |
| $A_F$ | percent | $98.6 \pm 1.9$ | $95.1 \pm 2.5$ |
| $P$ | day | 1.2749247646 (fixed) | |
| $e$ | – | 0 (fixed) | |
| $u_1$ | – | 0.072 (fixed) | |
| $u_2$ | – | 0.301 (fixed) | |

**Extended Data Table 2. Results of ATMO "2x PT" retrieval analyses.** Values listed are the posterior medians, with uncertainties giving the 1σ credible ranges. First row gives the results of the retrieval analysis performed for the dayside emission spectrum, with metallicity ([M/H]) fitted as a free parameter. Based on the retrieved metallicity for the dayside spectrum, the metallicity was held fixed to [M/H]=0.7 (i.e. 5x solar) for all other orbital phases. Last row gives the results of an additional retrieval performed as a check for phase $\phi$=0.95 with [M/H] treated as a free parameter.

| $\phi$ | $\kappa_{IR}$ (cm$^2$.g$^{-1}$) | $\gamma_1$ | $\Psi$ | [ M/H ] (dex) |
|---|---|---|---|---|
| **Dayside** | $-1.78^{+0.08}_{-0.26}$ | $1.76^{+0.16}_{-0.30}$ | $1.22^{+0.01}_{-0.01}$ | $0.76^{+0.30}_{-0.62}$ |
| **0.05** | $-2.99^{+1.19}_{-0.99}$ | $-0.52^{+2.26}_{-2.16}$ | $0.30^{+0.18}_{-0.15}$ | 0.7 (fixed) |
| **0.12** | $-2.09^{+0.91}_{-1.31}$ | $-1.48^{+2.09}_{-1.54}$ | $0.62^{+0.09}_{-0.14}$ | 0.7 (fixed) |
| **0.17** | $2.59^{+1.24}_{-1.36}$ | $-1.09^{+1.45}_{-1.65}$ | $0.73^{+0.06}_{-0.11}$ | 0.7 (fixed) |
| **0.23** | $-2.33^{+0.98}_{-1.83}$ | $-0.52^{+3.00}_{-2.03}$ | $0.64^{+0.11}_{-0.21}$ | 0.7 (fixed) |
| **0.28** | $-2.05^{+0.91}_{-1.82}$ | $-0.72^{+3.19}_{-1.64}$ | $0.73^{+0.09}_{-0.21}$ | 0.7 (fixed) |
| **0.32** | $-2.32^{+0.71}_{-0.45}$ | $3.28^{+0.43}_{-2.85}$ | $1.31^{+0.04}_{-0.05}$ | 0.7 (fixed) |
| **0.38** | $-2.17^{+0.47}_{-0.34}$ | $3.29^{+0.40}_{-1.48}$ | $1.30^{+0.02}_{-0.05}$ | 0.7 (fixed) |
| **0.43** | $-1.75^{+0.11}_{-0.14}$ | $1.64^{+0.09}_{-0.13}$ | $1.25^{+0.03}_{-0.03}$ | 0.7 (fixed) |
| **0.57** | $-2.04^{+0.34}_{-0.44}$ | $3.11^{+0.50}_{-1.27}$ | $1.27^{+0.02}_{-0.04}$ | 0.7 (fixed) |
| **0.62** | $-1.78^{+0.18}_{-0.16}$ | $1.79^{+0.12}_{-0.14}$ | $1.22^{+0.05}_{-0.04}$ | 0.7 (fixed) |
| **0.68** | $-2.77^{+1.17}_{-1.80}$ | $0.20^{+0.71}_{-0.49}$ | $1.13^{+0.19}_{-0.14}$ | 0.7 (fixed) |
| **0.72** | $-2.61^{+1.01}_{-1.32}$ | $-0.99^{+2.64}_{-1.66}$ | $0.40^{+0.18}_{-0.21}$ | 0.7 (fixed) |
| **0.78** | $-2.55^{+1.32}_{-1.72}$ | $-0.86^{+3.51}_{-1.75}$ | $0.70^{+0.08}_{-0.14}$ | 0.7 (fixed) |
| **0.82** | $-2.67^{+1.45}_{-1.62}$ | $-1.23^{+2.32}_{-1.65}$ | $0.69^{+0.07}_{-0.14}$ | 0.7 (fixed) |
| **0.88** | $-1.88^{+0.53}_{-0.62}$ | $-1.83^{+1.21}_{-1.10}$ | $0.71^{+0.06}_{-0.08}$ | 0.7 (fixed) |
| **0.95** | $-1.89^{+0.56}_{-0.56}$ | $-1.90^{+1.24}_{-1.13}$ | $0.70^{+0.06}_{-0.08}$ | 0.7 (fixed) |
| **0.95 (free [M/H])** | $-1.76^{+0.54}_{-0.92}$ | $-1.82^{+0.98}_{-1.30}$ | $0.70^{+0.06}_{-0.10}$ | $0.66^{+0.70}_{-1.02}$ |